\documentclass[10pt, a4paper]{article}
\usepackage[italian,british,UKenglish,USenglish,english,american]{babel}

\usepackage{amsmath}
\usepackage{helvet}
\usepackage{courier}
\usepackage{amssymb}
\usepackage{amsthm}
\usepackage[T1]{fontenc}
\usepackage[latin9]{inputenc}
\usepackage{makeidx}         
\usepackage{graphicx}        
\usepackage{multicol}        
\usepackage[bottom]{footmisc}

\newtheorem{definition}{Definition}
\newtheorem{theorem}{Theorem}
\newtheorem{proposition}{Proposition}

\newtheorem{corollary}{Corollary}
\usepackage{hyperref}
\hypersetup{colorlinks=true,
  linkcolor=blue,
  anchorcolor=blue,
  citecolor=red,
  urlcolor=brown,
  pdfauthor={E. Agliari, A. Barra, A. Galluzzi, F. Guerra, D. Tantari and F. Tavani}}

\providecommand{\be}{\begin{equation}}
  \providecommand{\ee}{\end{equation}}
\providecommand{\bea}{\begin{eqnarray}}
  \providecommand{\eea}{\end{eqnarray}}
\providecommand{\beas}{\begin{eqnarray*}}
  \providecommand{\eeas}{\end{eqnarray*}}

\providecommand{\beni}{\begin{equation*}}
  \providecommand{\eeni}{\end{equation*}}

\providecommand{\bw}{\begin{widetext}}
  \providecommand{\ew}{\end{widetext}}

\newcommand{\benumerate}{\begin{enumerate}}
\newcommand{\eenumerate}{\end{enumerate}}

\makeindex

\date{}

\begin{document}

\author{Elena Agliari$\;^{*}$ \; Adriano Barra$\;^{*}$ \; Andrea Galluzzi$\;^{**}$ \; Francesco Guerra$\;^{*}$ \; \\ Daniele Tantari$\;^{**}$ \; and \;  Flavia Tavani$\;^{***}$ \\
\footnotesize{$^{*}$Dipartimento di Fisica, Sapienza Universit\`{a} di Roma, Roma, Italy}\\
\footnotesize{$^{**}$Dipartimento di Matematica, Sapienza Universit\`{a} di Roma, Roma, Italy}\\
\footnotesize{$^{***}$Dipartimento di Scienze di base e applicate all'Ingegneria, Sapienza Universit\`{a} di Roma, Roma, Italy}}
\title{Meta-stable states in the hierarchical Dyson model drive parallel processing in the hierarchical Hopfield network}
\maketitle

\begin{abstract}
In this paper we introduce and investigate the statistical mechanics of hierarchical neural networks: First, we approach these systems \`a la Mattis, by thinking at the Dyson model as a single-pattern hierarchical neural network and we discuss the stability of different retrievable states as predicted by the related self-consistencies obtained from a mean-field bound and from a bound that bypasses the mean-field limitation. The latter is worked out by properly reabsorbing fluctuations of the magnetization related to higher levels of the hierarchy into effective fields for the lower levels. Remarkably, mixing Amit's ansatz technique (to select candidate retrievable states) with the interpolation procedure (to solve for the free energy of these states) we prove that   (due to gauge symmetry) the Dyson model accomplishes both serial and parallel processing.
\newline
One step forward, we extend this scenario toward multiple stored patterns by implementing the Hebb prescription for learning within the couplings. This results in an Hopfield-like networks constrained on a hierarchical topology, for which, restricting to the low storage regime (where the number of patterns grows at most logarithmical with the amount of neurons), we prove the existence of the thermodynamic limit for the free energy and we give an explicit expression of its mean field bound and of the related improved bound.
\newline
The resulting self-consistencies for the Mattis magnetizations (that act as order parameters) are studied and the stability of solutions is analyzed to get a picture of the overall retrieval capabilities of the system according to the mean field and to  the non-mean-field scenarios. Our main finding is that embedding the Hebbian rule on a hierarchical topology allows the network to accomplish both serial and parallel processing. By tuning the level of fast noise affecting it, or triggering the decay of the interactions with the distance among neurons, the system may switch from sequential retrieval to multitasking features and vice versa. However, as these multitasking capabilities are basically due to the vanishing ``dialogue'' between spins at long distance, such an effective penury of links strongly penalizes the network's capacity, which results bounded by the low storage.
\end{abstract}

\section*{Introduction}
Neural networks are such a fascinating field of science to attract an incredibly large variety of scientists, ranging from \emph{engineers} (mainly involved in electronics and robotics) \cite{1,2}, \emph{physicists} (mainly involved in statistical mechanics and stochastic processes) \cite{amit,hertz}, and \emph{mathematicians} (mainly working in learning algorithms and graph theory) \cite{peter,hinton} to \emph{(neuro) biologists} \cite{7,8} and \emph{(cognitive) psychologists} \cite{9,10}.

Tracing the genesis and evolution of neural networks back in time is very difficult, probably due to the broad meaning they have acquired along the years\footnote{Seminal ideas regarding automation are already in the works of Lee during the XIIX century, if not even back to Descartes, while more modern ideas regarding {\em spontaneous cognition}, can be attributed to A. Turing \cite{turing} and J. Von Neumann \cite{neumann} or to the join efforts of M. Minsky and S. Papert \cite{minchi}, just to cite a few.}: scientists closer to the robotics branch often refer to the W. McCulloch and W. Pitts model of perceptron \cite{culloch} (or the F. Rosenblatt version \cite{rosenblatt}), while researchers closer to the neurobiology branch adopt usually the D. Hebb work as a starting point \cite{hebb}.

On the other hand, scientists involved in statistical mechanics, that joined the community in relatively recent times (after a satisfactory picture of spin glasses was achieved \cite{MPV,dmitry1}, thus in the $'80$), usually refer to the seminal paper by Hopfield \cite{hopfield} or to the celebrated work by Amit, Gutfreund and Sompolinsky \cite{AGS}, where the statistical mechanical analysis of the Hopfield model is effectively carried out.

Confining ourselves within this last perspective and in a streamlined synthesis, the Hopfield model is a mean field model where neurons are mimicked by binary (Ising) spins, whose possible states represent firing or quiescence respectively \cite{amit,peter}, and which interact pairwise via the Hebb prescription. This model acts as the {\em harmonic oscillator} for serial processing: once the system is allowed to relax, it spontaneously retrieves one of the stored patterns (in suitably regions of the tunable parameters, e.g. low noise level and not-too-high storage load), pattern retrieval depending on, e.g. the initial state of the system. Recently, a generalization of this paradigm, i.e. the multitasking associative network \cite{PRL}, appeared as a candidate mean-field network  able to perform spontaneously parallel retrieval \cite{Sollich,ton1,ton2,Dantoni,jtb2,PRE}, that is to retrieve more patterns at once (without falling into spurious states) \cite{glassyNN,hotel}.
\newline
While these two networks perform in a crucial different way (serial versus parallel), they share the same mean-field statistical mechanics approximation: each neuron interacts with all the others it is linked to with the same strength, unaware of any underlying topology, namely independently of the actual pairwise {\em distance} among the neurons themselves. This limitation has always been considered as something to remove as soon as mathematical improvements of available techniques would allow.  Far from Artificial Intelligence, but exactly to this task (i.e. bypassing mean field limitations), a renewal interest is nowadays raised for hierarchical models, namely models where the closer the spins the stronger their links (see Fig. $1$). Starting from the pioneering Dyson work \cite{Dyson}, where the hierarchical ferromagnet was introduced and its phase transition (splitting an ergodic region from a ferromagnetic one) rigorously proven, recently its extensions to spin-glasses have also been investigated \cite{REM}. Although an analytical solution is still not available, giant step forward toward a deep comprehension of the hierarchical statistical mechanics have been obtained \cite{Castellana1,DH,Franz,leuzzi1,leuzzi2,cecilia1,cecilia2,mukamel}.
\newline
In this paper we aim to analyze in details the hierarchical neural networks, and to this task we start by considering the statistical mechanics of the Dyson model from a novel perspective:  we investigate its metastabilities.

In Section One we deal with Dyson's model: once fundamental definitions have been introduced, in the first subsection we prove the existence of the thermodynamic limit of its related free energy within the spirit of the classical Guerra-Toninelli scheme \cite{GT} to the case. The following subsection is dedicated to a mean-field picture: we mix Amit's ansatz technique \cite{amit} with our interpolation schemes \cite{broken}; the resulting technique allows to think at the Dyson model as a single-pattern associative network (as the Curie-Weiss plays in mean-field counterpart thanks to the Mattis gauge \cite{peter}): a satisfactory picture of its related thermodynamic and retrieval capabilities is obtained and discussed: Remarkably, the intrinsic richness of  (effectively) possible states in Dyson model drives the system from serial processing to parallel processing and yields the breakdown of the self-average for the order parameters. Note that {\em parallel processing} may appear strange for a {\em one-pattern} neural network, however due to gauge-symmetry, the stored patterns are actually two, thus if one-half of the network spins retrieve the original pattern and the other half its gauged version, this can be seen as a multitasking feature, as it will become obvious when investigating the hierarchical Hopfield model.
\newline
The next subsection $1.3$ traces the same line of $1.2$ but bypassing mean-field limitation: despite we are not able to completely solve the statistical mechanics of this model yet, thanks to a new interpolation scheme (developed in \cite{DH}), we are able to account (partially) for order parameter's fluctuations level by level (of the hierarchy). The idea is to leverage the hierarchical structure of the model in order to account for such fluctuations: the latter are reabsorbed -again recursively, i.e. level by level- into an effective Hamiltonian for the underlying block spins whose thermodynamics remains still solvable, thus improving the mean-field result. The difference between these two scenarios lies only in a different critical noise for ergodicity breaking, but serial and parallel retrieval capabilities (namely the existence of pure and meta-stable states) are preserved in both cases.
Following the strand paved for the Dyson analysis, in Section Two we introduce the real hierarchical neural network, namely a hierarchical network with Hebbian couplings or, equivalently, the Hopfield model on a hierarchical setting. This system is studied in the low storage regime, that is where the amount of stored patterns scales at most logarithmical with the amount of neurons the network is built with.
For this model we prove at first the existence of its free energy's thermodynamic limit (subsection $2.1$), then we move toward a mean-field scenario ($2.2$), further we investigate the non-mean-field one ($2.3$). In both cases, the model has an extremely rich phase diagram, where beyond standard serial retrieval (which is accomplished too), a number of parallel states suddenly appears by properly tuning the level of (fast) noise affecting the network.
A discussion of these states and their stability analysis is included still in Sec.$2$, while a discussion regarding network's capacity can be found in the following conclusions, which close the paper.

\section{Analysis of the Dyson hierarchical model}

The Dyson Hierarchical Model (DHM) is a system composed -at the microscopic level- by $2^{k+1}$ Ising spins $S_i=\pm 1$, with $i=1,...,2^{k+1}$ embedded in a hierarchical topology. The Hamiltonian capturing the model is recursively introduced by the following
\begin{figure}[tb] \begin{center}
\includegraphics[width=.42\textwidth]{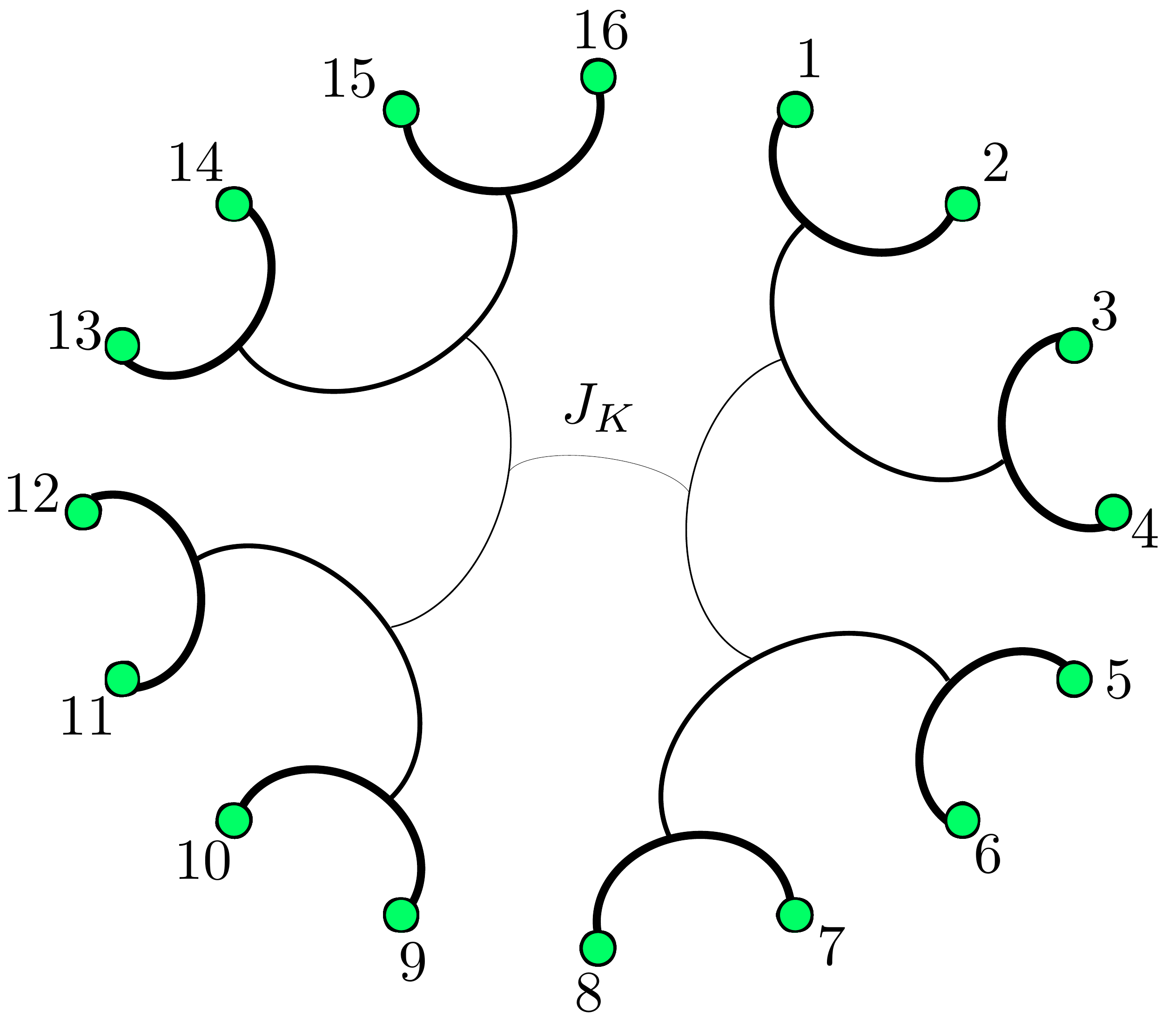}
\;\;
\includegraphics[width=.45\textwidth]{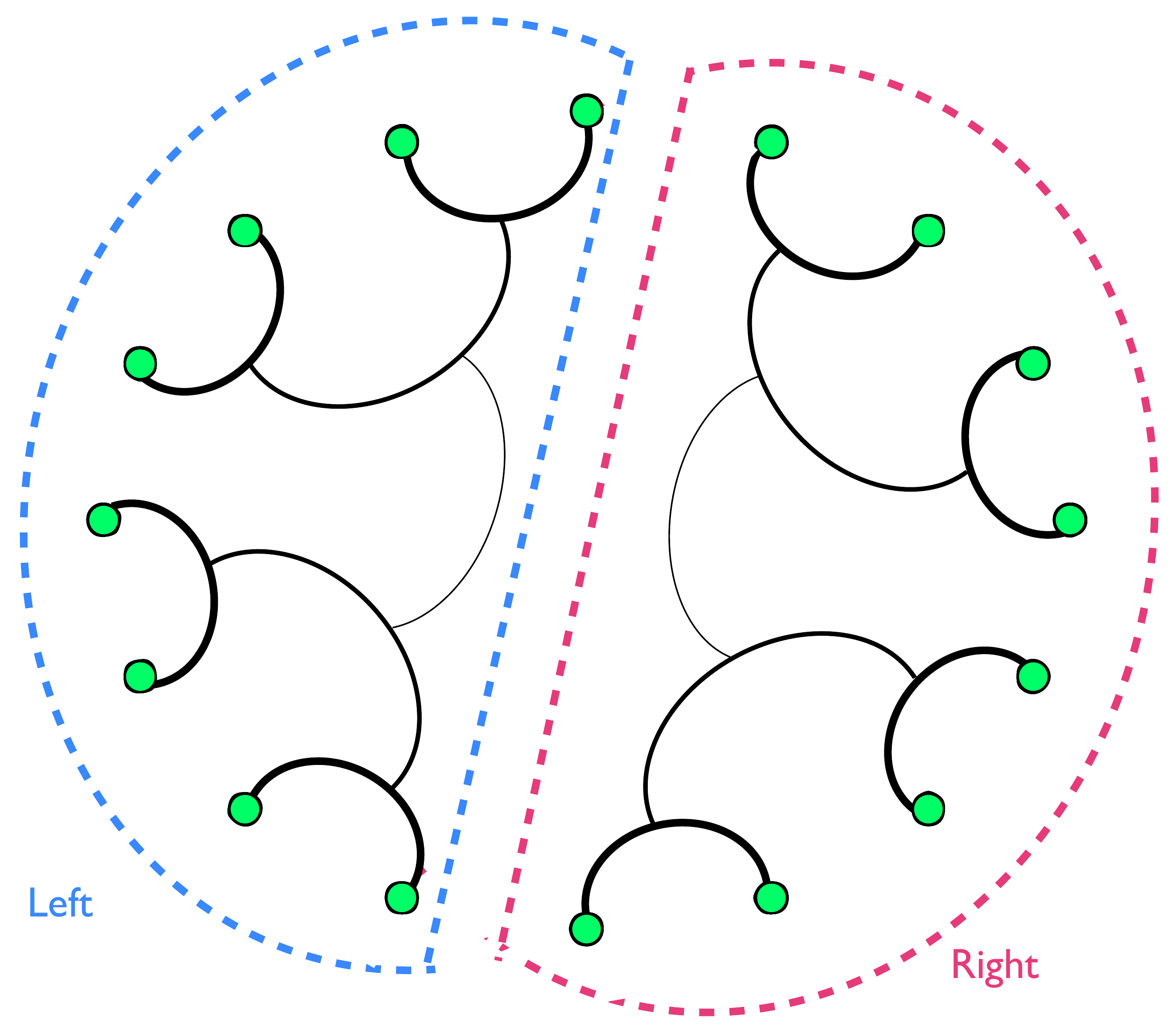}
\caption{\label{fig:esempio} Schematic representation of the hierarchical topology where the associative network insists. Green spots represent Ising neurons ($N=16$ in this shapshot) while links are drawn with different thickness mimicking various interaction strengths: The thicker the line, the stronger the link.
}
\end{center}
\end{figure}
\begin{definition}
The Hamiltonian of Dyson's Hierarchical Model (DHM) is defined by
\be\label{disonne}
H_{k+1}(\vec{S}|J,\sigma)=H_k(\vec{S_1})+H_k(\vec{S_2})-\frac{J}{2^{2\sigma(k+1)}}\sum_{i<j=1}^{2^{k+1}}S_iS_j,
\ee
where $J>0$ and $\sigma \in (1/2,1)$ are numbers tuning the interaction strength. Clearly $\vec{S_1}\equiv \{ S_i \}_{1 \leq i \leq 2^k}$,  $\vec{S_2}\equiv \{ S_j \}_{2^k+1 \leq j \leq 2^{k+1}}$ and $H_0[S]=0$.
\end{definition}
Thus, in this model, $\sigma$ triggers the decay of the interaction with the distance among spins, while $J$ uniformly rules the overall intensity of the couplings.
\newline
Note that this model is explicitly a non-mean-field model as the distance $d_{i,j}$  between two spins $i,j$ ranges in $0$ and $k$ (see Fig.$1$). Indeed, it is possible to re-write the Hamiltonian (\ref{disonne}) in terms of the $d_{i,j}$ as
\begin{eqnarray}
H_{k}[\{S_1...S_{2^k}\}]&=&-\sum_{i<j}S_i S_j J_{ij}\\
J_{ij}&=&\sum_{l=d_{i,j}}^{k} \left ( \frac{J}{2^{2\sigma l}} \right)=J(d_{ij},k,\sigma,J)=J\frac{4^{\sigma -d_{i,j}  \sigma }-4^{-k \sigma }}{4^{\sigma }-1}.
\end{eqnarray}
Once the Hamiltonian is given (and in this paper we will refer mainly to the form (\ref{disonne})), it is possible to introduce the partition function $Z_{k+1}(\beta,J,\sigma)$ at finite volume $k+1$ as
\be
Z_{k+1}(\beta,J,\sigma) = \sum_{\sigma}^{2^{{k+1}}}\exp\left[-\beta H_{k+1}(\vec{S}|J,\sigma)\right],
\ee
and the related free energy $f_{k+1}(\beta,J,\sigma)$, namely the intensive logarithm of the partition function, as
\be
f_{k+1}(\beta,J,\sigma)=\frac{1}{2^{k+1}}\log\sum_{\vec{S}}\exp \left[ -\beta H_{k+1}(\vec{S})+h \sum_{i=1}^{2^{k+1}}S_i \right].
\ee
We are interested in an explicit expression of the infinite volume limit of the intensive free energy, defined as
$$
f(\beta,J,\sigma)= \lim_{k\to\infty}f_{k+1}(\beta,J,\sigma),
$$
in terms of suitably introduced magnetizations $m$, that act as order parameters for the theory because -in order to satisfy thermodynamic prescriptions- we want to find the free energy minima\footnote{Note that as the free energy, strictly speaking, is $f(\beta)=-\alpha(\beta)$, we actually look for maxima trough the paper.} w.r.t. these order parameters. To this task we introduce the global magnetization $m$, defined as the limit $m=\lim_{k\to\infty}m_{k+1}$ where
\be
m_{k+1}=\frac{1}{2^{k+1}}\sum_i^{2^{k+1}}S_i,
\ee
and, recursively and with a little abuse of notation,
the $k$ magnetizations $m_a,...,m_k$ level by level (over $k$ levels and starting to defined them from the largest bulk), as the same $k\to\infty$ limit of the following quantities (we write explicitly only the two upper magnetizations related to the two main clusters the system reduces to whenever $J_K\to0$ -see Fig.$1$-):
\be
m_{left}=\frac{1}{2^{k}}\sum_{i=1}^{2^{k}}S_i, \ \ \ m_{right}=\frac{1}{2^{k}}\sum_{i=2^k+1}^{2^{k+1}}S_i.
\ee
As a last point, thermodynamical averages will be denoted by the brackets $\langle \cdot \rangle$, such that
\be
\langle m_{k+1}(\beta,J,\sigma) \rangle = \frac{\sum_{\sigma} m_{k+1}e^{-\beta H_{k+1}(\vec{S}|J,\sigma)}}{Z_{k+1}(\beta,J,\sigma)},
\ee
and clearly $\langle m(\beta,J,\sigma) \rangle = \lim_{k \to \infty} \langle m_{k+1}(\beta,J,\sigma) \rangle$.

\subsection{The thermodynamic limit}

Argument of this Section is a proof of the existence of the thermodynamic limit for the free energy of the DHM: despite this result has been already achieved a long time ago by Gallavotti and Miracle-Sole \cite{Gallavotti}, we exploit here a different interpolating scheme with the pedagogical aim of highlighting the technique more than the result itself as it will then be used to prove the existence of the thermodynamic limit for the hierarchical Hopfield network. The main idea is that, since the interaction is ferromagnetic, the free energy is monotone in $k$, with the introduction of new levels of positive interactions.
\begin{theorem}
The thermodynamic limit of the DHM free energy does exist and we call
 $$\lim_{k\rightarrow\infty}f_{k+1}(\beta,J,\sigma)= f(\beta,J,\sigma).$$
\end{theorem}

To prove this statement let us introduce a real scalar parameter $t\in[0,1]$ and the following interpolating function
\be
\Phi_{k+1,t}(\beta)=\frac{1}{2^{k+1}}\log\sum_{\vec{S}}\exp(\beta(H_k(\vec{S_1})+H_k(\vec{S_2})+\frac{tJ}{2}2^{(k+1)}2^{(k+1)(1-2\sigma)}m_{k+1}^2(\vec{S})),
\ee

with  $m_{k+1}=\frac{1}{2^{k+1}}\sum_{l=1}^{2^{k+1}}S_l$, such that
\begin{eqnarray}
\Phi_{k+1,1}&=&f_{k+1},\\
\Phi_{k+1,0}&=&f_k,
\end{eqnarray}
and
\be
0\leq\frac{d\Phi_{k+1,t}}{dt}=\left\langle\beta\frac{1}{2^{k+1}}\frac{2^{(k+1)}2^{(k+1)(1-2\sigma)}J}{2}m_{k+1}^2(\vec{S})\right\rangle_t\leq\frac{\beta J 2^{(k+1)(1-2\sigma)}}{2}.
\ee

Since

$$\Phi_{k+1,1}(h)=\Phi_{k+1,0}(h)+\int_{0}^1 \frac{d\Phi_{k+1,t}}{dt}dt,$$

$f_{k+1}\geq f_k$ (the sequence is non-decreasing), thus
\be
f_{k+1}(\beta,J,\sigma)\leq f_k(\beta,J,\sigma)+\frac{\beta J}{2}2^{(k+1)(1-2\sigma)}.
\ee
Iterating this argument over the levels we obtain
\be
f_{k+1}(\beta,J,\sigma)\leq f_0(\beta,J,\sigma)+\frac{\beta J}{2}\sum_{l=1}^{k+1}2^{l(1-2\sigma)}.
\ee

In the limit of $k\rightarrow\infty$
\be
f\leq f_0+\frac{\beta J}{2}\sum_{l=1}^{\infty}2^{l(1-2\sigma)}.
\ee
The series on the right of the above inequality converges,  since $\sigma >\frac{1}{2}$, hence
\be
f(\beta,J,\sigma)\leq f_0(\beta,J,\sigma)+\frac{\beta J}{2}\frac{1}{1 - 2^{(2\sigma-1)}}.
\ee

The sequence $f_k(\beta,J,\sigma)$ is bounded and non-decreasing, so it admits a well defined limit for $k\rightarrow\infty$.


\subsection{The mean-field scenario}

Plan for this Section is to turn around classical results \cite{Dyson,Gallavotti,Sinai} to investigate meta-stabilities in the Dyson model at the mean-field level. To this task two schemes must be merged: we start following \cite{DH} for building an interpolating iterative scheme that returns the mean-field free energy in terms of a bound, then we implement the Amit method of ansatz to evaluate -within the free energy landscape obtained by this interpolation- the stability and thermodynamic importance of two test-states: the (standard) ferromagnetic state (with all the spin aligned, hence $m_{left}=m_{right}$) and the simplest meta-stable state, namely a state where all the left spins (that is the first $1,...,2^k$ spins) are aligned each other and opposite to the right spins (that is the remaining $2^k+1,...,2^{k+1}$ spins), which -in turn- are aligned each other too (hence $m_{left}=-m_{right}$). Operatively, we state the next
\begin{definition}
Once considered a real scalar parameter $t\in[0,1]$, we introduce the following interpolating Hamiltonian
\be
H_{k+1,t}(\vec{S})=-\frac{Jt}{2^{2\sigma(k+1)}}\sum_{i>j=1}^{2^{k+1}}S_iS_j-(1-t)mJ2^{(k+1)(1-2\sigma)}\sum_{i=1}^{2^{k+1}}S_i+H_k(\vec{S_1})+H_k(\vec{S_2}),
\ee
such that for $t=1$ the original system is recovered, while at $t=0$ the two body interaction is replaced by an effective but tractable one-body term. The possible presence of an external magnetic field can be accounted simply by adding to the Hamiltonian a term $\propto h\sum_i^{2^{k+1}}\sigma_i$, with $h \in \mathcal{R}$.
\end{definition}
This prescription allows defining an extended partition function as
\be\label{etti}
Z_{k+1,t}(h,\beta,J,\sigma)=\sum_{\vec{S}}\exp \{ -\beta [ H_{k+1,t}(\vec{S})+h\sum_{i=1}^{2^{k+1}}S_i ] \},
\ee
where the subscript $t$ stresses its interpolative nature, and, analogously,
\be
\Phi_{k+1,t}(h,\beta,J,\sigma)=\frac{1}{2^{k+1}}\log Z_{k+1,t}(h,\beta,J,\sigma).
\ee
Since
\be \Phi_{k+1,0}(h,\beta,J,\sigma)=\Phi_{k,1}(h+ mJ2^{(k+1)(1-2\sigma)},\beta,J,\sigma),\ee
as shown in \cite{DH}, (discarding the dependence of $\Phi$ by $\beta,\ J,\ \sigma$ for simplicity) through a long but straightforward calculation, we arrive to
\begin{eqnarray}
\Phi_{k+1,1}(h)&=&\Phi_{k+1,0}(h)-\frac{\beta J}{2}(2^{(k+1)(1-2\sigma)}m^2+2^{-2(k+1)\sigma})+\frac{\beta J}{2}2^{(k+1)(1-2\sigma)}\left\langle (m_{k+1}(\vec{S})-m)^2\right\rangle_t \nonumber \\
&\geq&\Phi_{k,1}(h+ Jm2^{(k+1)(1-2\sigma)})-\frac{\beta J}{2}(2^{(k+1)(1-2\sigma)}m^2+2^{-2(k+1)\sigma}).
\end{eqnarray}
Note that, in the last passage, we neglected level by level the source of order parameter's fluctuations $\left\langle (m_{k+1}(\vec{S})-m)^2\right\rangle_t$ -which is positive definite- thus we obtained a bound for the free energy.
\newline
For the seek of simplicity we extended the meaning of the brackets to account also for the interpolating structure coded in the Boltzmannfaktor of eq.(\ref{etti}), by adding to them a subscript $t$, namely $\langle \cdot \rangle \to \langle \cdot \rangle_t$.

In order to start investigating non-standard stabilities, note further that $\Phi_{k+1,0}(h)=\Phi_{k,1}(h+ mJ2^{(k+1)(1-2\sigma)})$ but in principle we can have also two different contributions from the two groups of $2^k$ spins ({\em left} and {\em right}) thus we should write more generally
\be
\Phi_{k+1,0}(h)=\frac 1 2 \left[ \Phi^1_{k,1}(h+ mJ2^{(k+1)(1-2\sigma)})+\Phi^2_{k,1}(h+ mJ2^{(k+1)(1-2\sigma)})\right ].
\ee
Now let us assume the Amit perspective \cite{amit} and suppose that these two subsystems have different magnetizazions  $m_{left}=m_1$ and $m_{right}=m_2$ (equal in modulus but opposite in sign, i.e. $m_1=-m_2$): this observation implies that, starting from the $k$-th level, we can iterate the interpolating procedure in parallel on the two clusters using respectively $m_1$ and $m_2$ as trial parameters. Via this route we obtain
\begin{eqnarray}
\Phi_{k+1,1}(h)&\geq& \frac{1}{2}\Phi_{0,1} \left \{ h+ J \left[ \sum_{l=1}^{k}2^{l(1-2\sigma)}m_1+2^{(k+1)(1-2\sigma)}m \right] \right \}\nonumber\\
&+&\frac{1}{2}\Phi_{0,1} \left \{ h+ J \left[ \sum_{l=1}^{k+1}2^{l(1-2\sigma)}m_2+2^{(k+1)(1-2\sigma)}m \right]  \right \}\nonumber\\
&-&\frac{\beta J}{2} \left[ 2^{(k+1)(1-2\sigma)}m^2+\sum_{l=1}^{k+1}2^{-2l\sigma} \right]-\frac{\beta J}{2}\sum_{l=1}^{k}2^{l(1-2\sigma)}\left(\frac{m_1^2+m_2^2}{2}\right),
\end{eqnarray}
that is
\bea\label{mfdys}
f_{k+1}(h,\beta,J,\sigma)&\geq& \log 2 +\frac{1}{2} \left \{ \log\cosh \left [ \beta h+\beta J \left( m_1\sum_{l=1}^k 2^{l(1-2\sigma)}+2^{(k+1)(1-2\sigma)}m \right)  \right ] \right \}+\nonumber\\
&+&\frac{1}{2} \left \{ \log\cosh \left[ \beta h+\beta J \left( m_2\sum_{l=1}^k 2^{l(1-2\sigma)}+2^{(k+1)(1-2\sigma)}m \right )\right ] \right \}+
\nonumber\\
&-&\frac{\beta J}{2} \left[ 2^{(k+1)(1-2\sigma)}m^2+\sum_{l=1}^{k+1}2^{-2l\sigma} \right] -\frac{\beta J}{2}\sum_{l=1}^{k}2^{l(1-2\sigma)} \left ( \frac{m_1^2+m_2^2}{2} \right) \nonumber \\
&=& f(k,m,m_1,m_2|h,\beta,J,\sigma).
\eea
\noindent Therefore, we have that $f_{k+1}(h,\beta,J,\sigma)\geq \sup_{m,m_1,m_2}f(k,m,m_1,m_2|(h,\beta,J,\sigma)$ and we need to evaluate the optimal order parameters in order to have the best free energy estimate.
\newline
Taking the derivatives of the free energy with respect to $m$, $m_1$ and $m_2$ we obtain the self consistent equations holding at the extremal points of $f(k,m,m_1,m_2|h,\beta,J,\sigma)$, which read as
\beas
\left\{
\begin{array}{l}
m_1=\tanh \left [ \beta h+\beta J \left( m_{1}\sum_{l=1}^k 2^{l(1-2\sigma)}+2^{(k+1)(1-2\sigma)}m \right) \right] ,\\\\
m_2=\tanh \left [ \beta h+\beta J \left ( m_{2}\sum_{l=1}^k 2^{l(1-2\sigma)}+2^{(k+1)(1-2\sigma)}m \right )\right],\\\\
m=\frac{m_1+m_2}{2},
\end{array}
\right.
\eeas
where the third equation is only a linear combination of $m_1$ and $m_2$ and it simply states that the global magnetization is the average of the ones of the two main clusters.

It is easy to see that, at zero external field $h=0$, the Pure solution $m_1=m_2=m=m_P$, where the whole system has a non zero magnetization, and the Antiparallel (meta-stable) one $m_A=m_1=-m_2$ and $m=0$, where the system has two clusters with opposite magnetizations and no global magnetization, both exist.

Clearly, according to the value of the temperature, we can have a paramagnetic solution ($m_P=m_A=0$), or two gauge symmetric solutions for each of the two possible states ($\pm m_P,\pm m_A$); we therefore need to analyze the stability of these solutions, checking if they are maxima or minima of $f(h=0,\beta,J,\sigma)$.
\newline
Obtaining an explicit expression for the second derivatives to build the Hessian $H(m_1,m_2)$ of $f(h=0,\beta,J,\sigma)$ is rather lengthy, yet it is easy to see that the entries of Hessian actually depends on $m_1^2$ and on $m_2^2$ only, namely they are independent of the sign of the two magnetizations $m_1,m_2$. This means that, as the paramagnetic solution becomes unstable, both the pure (i.e. $m_{left}=m_{right}$) and antiparallel (mixture, i.e.  $m_{left}= -m_{right}$) solutions become stable (this ensures the possibility to take the thermodynamic limit in ($\ref{mfdys}$) and sheds lights on breaking of standard self-averaging \cite{Sinai}).
\newline
In this case we get the following
\begin{theorem}
The mean-field bound for the DHM free energy associated to the meta-stable state reads as
\bea
f(h,\beta,J,\sigma)\geq \sup_{m_1,m_2,m}&& \lim_{k\to \infty} f(k,m,m_1,m_2)\nonumber\\
= \sup_{m_1,m_2}&& \log 2 +\frac{1}{2}\log\cosh(\beta h+\beta J C_{2\sigma-1} m_1)\nonumber\\
&+&\frac{1}{2}\log\cosh((\beta h+\beta J C_{2\sigma-1}m_2)-\frac{\beta J C_{2\sigma}}{2}-\frac{\beta J  C_{2\sigma-1}}{2}(\frac{m_1^2+m_2^2}{2}),
\eea
\noindent where $C_{y}=\frac{2^{-y}}{1-2^{-y}}$ and the trial parameters $\ m_1,\ m_2$ fulfill the self-consistencies \ref{selfo} and $m$ is its symmetric linear combination.
\newline
The mean field bound for the DHM free energy associated to the ferromagnetic state can be obtained again simply by identifying $m_1=m_2=m$ and reads as
\bea
f(h,\beta,J,\sigma)\geq \sup_{m}\left[ \log 2 +\log\cosh(  \beta h+\beta J C_{2\sigma-1} m)-\frac{\beta J C_{2\sigma}}{2}-\frac{\beta J  C_{2\sigma-1}}{2}m^2\right]
\eea
whose self-consistencies can be found in \cite{DH}.
\end{theorem}
In the thermodynamic limit, the last level of interaction (the largest in number of links but the weakest as for their intensity), that would tend to keep $m_1$ and $m_2$ aligned, vanishes. Thus the system effectively behaves just as the sum of two non interacting subsystems with independent magnetizations satisfying the following
\begin{proposition}
The mixture state of the DHM has two  independent order parameters, one for each larger cluster, whose self-consistencies read as
\be\label{selfo}
m_{1,2}=\tanh\left(\beta h+\beta J C_{2\sigma-1} m_{1,2}\right).
\ee
\end{proposition}
One step forward, if we want to find out the critical value  $\beta_c$ that breaks ergodicity, we can expand them for $k\rightarrow\infty$, and for $h=0$, hence obtaining
\beas
\left \{
\begin{array}{l}
m_1\sim \beta Jm_1\frac{2^{1-2\sigma}}{1-2^{1-2\sigma}}+\mathcal{O}(m_1^3),\\\\
m_2\sim \beta Jm_2\frac{2^{1-2\sigma}}{1-2^{1-2\sigma}}+\mathcal{O}(m_2^3),
\end{array}
\right \}
\eeas
such that we can write the next
\begin{corollary}
Mean-field criticality in the DHM has the classical critical exponent one half and critical temperature $\beta_c^{MF}$ given by
\be
\beta_c^{MF}=\frac{1-2^{1-2\sigma}}{J2^{1-2\sigma}}.
\ee
It is worth noticing however that the mean-field picture does not hold in this hierarchical setting.
\end{corollary}
One may still debate however that, while the intensity of the upper links is negligible, it  may still collapse the state of one cluster to the other (thus destroying  metastability), as for instance happens when we use a vanishing external field in a critical mean-field ferromagnet to select the phase by hand. In the appendix A we give a detailed explanation, and a rigorous proof, that this is not the case here: The DHM has links too {\em evanescent} to drive all the spins to converge always to the same sign and mixture states are preserved.
\begin{figure}[tb] \begin{center}
\includegraphics[width=.85\textwidth]{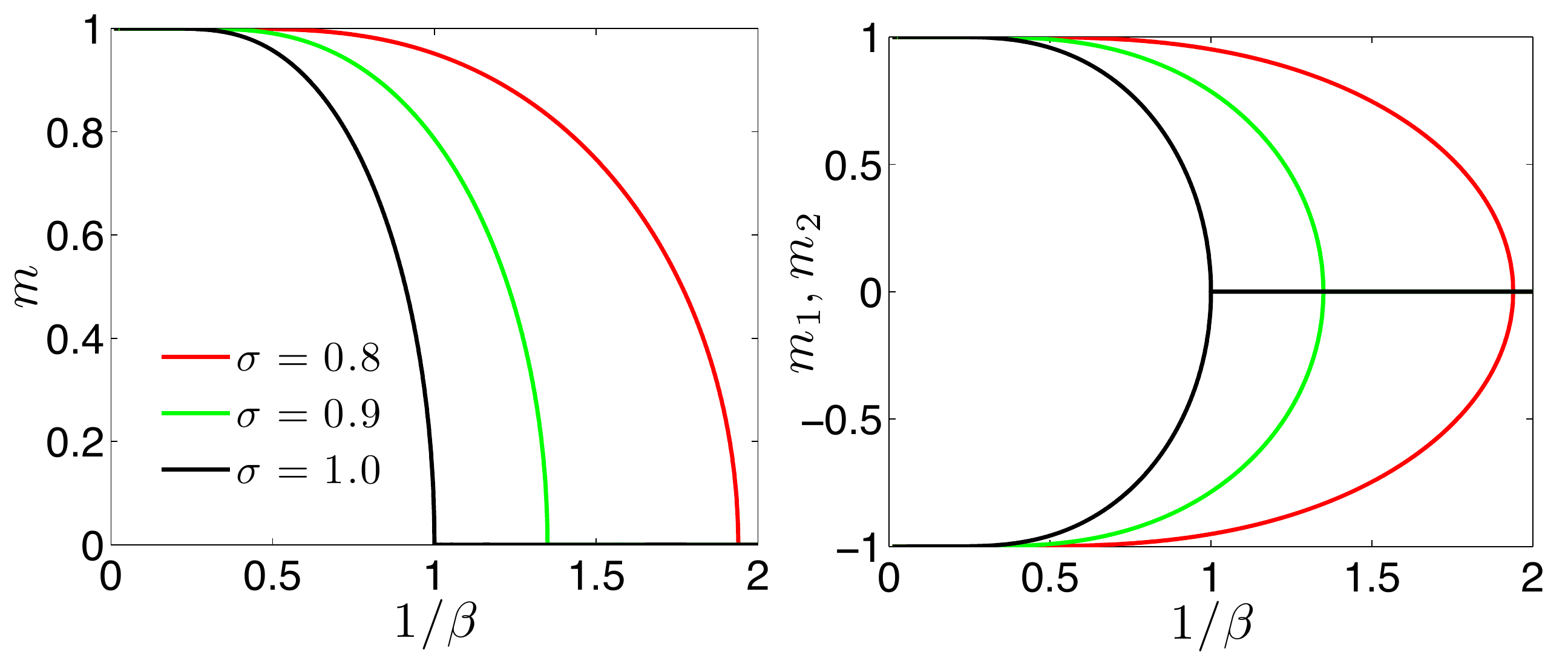}
\caption{Behavior of the magnetizations for the Dyson model within the non-mean-field scenario. Left panel: Pure state (serial processing). Right panel: Mixture state (parallel processing). Note that the difference in energy among the pure state and the mixture state scales as $\Delta E \propto 2^{(k+1)\cdot(1-2\sigma)}$, thus -in the thermodynamic limit- the parallel state becomes effectively stable (see Fig.$3$ too), but do not alter the thermodynamical picture of ferromagnetism.}
\end{center}
\end{figure}

\subsection{The not-mean-field scenario}

Scope of the present Section is to bypass mean-field limitations and show that the outlined scenario is robust even beyond the mean-field picture. We stress that we do not have a rigorous solution of the free energy, but rather a more stringent (with respect to the mean-field counterpart) analytical bound supported by extensive numerical simulations. In particular, we exploit the interpolative technology introduced in \cite{DH} to take into account (at least a) part of the fluctuations of the order parameters (thus improving the previous description) as, in models beyond mean-field, the magnetization is no longer self-averaging and its fluctuations can not be neglected. It is indeed the proliferation of these meta-stable states that avoids the collapse of the order parameter probability distribution on a Dirac delta and breaks self-averaging.
\newline
Let us start investigating the improved bound with the following
\begin{definition}
Once introduced two suitable real parameters $t,\ x$, the interpolating Hamiltonian that we are going to consider to bypass the mean-field bound has the form
\be
H_{k+1,t}(\vec{S})=-tu(\vec{S})-(1-t)v(\vec{S})+H_k(\vec{S_1})+H_k(\vec{S_2}),
\ee
with
\begin{eqnarray} \nonumber
u(\vec{S})&=&\frac{J}{2^{2\sigma(k+1)}}\sum_{i>j=1}^{2^{(k+1)}}S_iS_j+\frac{xJ}{2\cdot 2^{2\sigma(k+1)}}\sum_{i,j=1}^{2^{k+1}}(S_i-m)(S_j-m),\\ \nonumber
v(\vec{S})&=&\frac{J(1+x)}{2\cdot 2^{2\sigma(k+1)}} \left [ \sum_{i,j=1}^{2^k}(S_i-m)(S_j-m)+\sum_{i,j=2^k+1}^{2^{k+1}}(S_i-m)(S_j-m) \right] +mJ2^{(k+1)(1-2\sigma)}\sum_{i=1}^{2^{k+1}}S_i,
\end{eqnarray}
where $x\geq 0$ accounts for fluctuation resorption and $0\leq t\leq 1$ plays as before.
\end{definition}
The associated partition function and free energy are, respectively,
\begin{eqnarray}
Z_{k+1,t}(x,h)&=&\sum_{\vec{S}}\exp \left \{  -\beta \left[ H_{k+1,t}(\vec{S})+h\sum_{i=1}^{2^{k+1}}S_i \right] \right \}  ,\\
\Phi_{k+1,t}(x,h)&=&\frac{1}{2^{k+1}}\log Z_{k+1,t}(x,h).
\end{eqnarray}
The procedure that yields to the non-mean-field bound for the free energy permits to obtain (see \cite{DH}) the following expression for the pure ferromagnetic case (where again we omitted the dependence by $\beta,\ J,\ \sigma$ for simplicity)
\be
f_{k+1}(h,\beta,J,\sigma)\geq \Phi_{k,1} (\frac{1}{2^{2\sigma}},h+ m2^{(k+1)(1-2\sigma)})-\frac{\beta J}{2}(2^{(k+1)(1-2\sigma)}m^2+2^{-2\sigma(k+1)}).\label{NMFbound}
\ee
However, as shown for the previous bound, let us now suppose that the system is split in two parts, with two different magnetizations $m_{left}=m_1$ and $m_{right}=m_2$: resuming the same lines of reasoning of the previous Section, we obtain
\be
\Phi_{k,1}(\frac{1}{2^{2\sigma}},h+ m2^{(k+1)(1-2\sigma)})=\frac{1}{2}
\Phi^1_{k,1}(\frac{1}{2^{2\sigma}},h+ m 2^{(k+1)(1-2\sigma)})+\frac{1}{2} \Phi^2_{k,1}(\frac{1}{2^{2\sigma}},h+ m2^{(k+1)(1-2\sigma)}).
\ee
From this point we can iterate the previous scheme point by point up to the last level of the hierarchy using as trial order parameter $m_{1,2}$ for $\Phi^{1,2}$, respectively. As a consequence, formula $(\ref{NMFbound})$, derived within the ansatz of a pure ferromagnetic state, is generalized by the following expression
\beas
f_{k+1}(h,\beta,J,\sigma)\geq \frac{1}{2}\Phi_{0,1}(\sum_{l=1}^{k+1}2^{-2l\sigma},h+J m_1\sum_{l=1}^{k}2^{l(1-2\sigma)}+ mJ2^{(k+1)(1-2\sigma)})+\qquad\qquad\qquad\qquad\\+\frac{1}{2}\Phi_{0,1}(\sum_{l=1}^{k+1}2^{-2l\sigma},h+J m_2\sum_{l=1}^{k}2^{l(1-2\sigma)}+ mJ2^{(k+1)(1-2\sigma)})+\qquad\qquad\qquad\qquad\\-\frac{\beta J}{2}\sum_{l=1}^k2^{l(1-2\sigma)}(\frac{m_1^2+m_2^2}{2})-\frac{\beta J}{2}\sum_{l=1}^{k+1}2^{-2l\sigma}-\frac{\beta J}{2}2^{(k+1)(1-2\sigma)}m^2.\qquad\qquad\qquad\qquad
\eeas
An explicit representation for $\Phi_{0,1}$ reads as
\begin{eqnarray}
&&\Phi_{0,1}(\sum_{l=1}^{k+1}2^{-2l\sigma},h+ Jm_1\sum_{l=1}^{k}2^{l(1-2\sigma)}+ mJ2^{(k+1)(1-2\sigma)})= \ln 2 + \\ &&+\frac{\beta J}{2}(1+m_1^2)\sum_{l=1}^{k+1}2^{-2l\sigma}+\log\cosh \left \{\beta h + \beta mJ2^{(k+1)(1-2\sigma)}+\beta m_1J \left [\sum_{l=1}^{k}2^{l(1-2\sigma)}-\sum_{l=1}^{k+1}2^{-2l\sigma} \right]  \right \}\qquad\qquad\qquad\qquad\qquad\qquad\qquad,
\end{eqnarray}
in such a way that
\begin{eqnarray}\nonumber
f_{k+1}\geq \log 2 &+&\frac{1}{2}\log\cosh \left \{ \beta h+\beta mJ  2^{(k+1)(1-2\sigma)}+\beta m_1J \left [\sum_{l=1}^{k}2^{l(1-2\sigma)}-\sum_{l=1}^{k+1}2^{-2l\sigma} \right]  \right \}+ \\ \nonumber &+&\frac{1}{2}\log\cosh \left \{\beta h+\beta mJ 2^{(k+1)(1-2\sigma)} +\beta m_2J \left[ \sum_{l=1}^{k}2^{l(1-2\sigma)}-\sum_{l=1}^{k+1}2^{-2l\sigma} \right]  \right \}
+\\ \nonumber &-&\frac{\beta J}{2} \left[ \sum_{l=1}^{k}2^{l(1-2\sigma)}-\sum_{l=1}^{k+1}2^{-2l\sigma} \right] \left( \frac{m_1^2+m_2^2}{2} \right )+
\\&-&\frac{\beta J}{2}2^{(k+1)(1-2\sigma)}m^2.
\end{eqnarray}
Summarizing, in the thermodynamic limit one has the following
\begin{theorem}
The non-mean-field bound for the DHM's free energy associated to the mixture state reads as
\begin{eqnarray}
f(h,\beta,J,\sigma)&\geq&  \sup_{m_1, m_2}\ \
\Big \{ \log 2+\frac{1}{2}\log\cosh \left[ \beta h+\beta m_1J(C_{2\sigma-1}-C_{2\sigma}) \right ]+\\ \nonumber
&+&\frac{1}{2}\log\cosh\left[ \beta h+\beta m_2J(C_{2\sigma-1}-C_{2\sigma}) \right]
-\frac{\beta J}{2}(C_{2\sigma-1}-C_{2\sigma}) \left (\frac{m_1^2+m_2^2}{2} \right) \Big \},
\end{eqnarray}
where $C_y=\frac{2^{-y}}{1-2^{-y}}$, and the trial parameters $\ m_1,\ m_2$ respect the self-consistencies that we will outline in Proposition $2$. If we assume that the system lives within a pure state, identifying then $m_1=m_2=m$, we find again the non-mean-field bound shown in \cite{DH}, that is
\be
f(h,\beta,J,\sigma)\geq  \sup_{m} \left \{  \log 2+\log\cosh \left[ \beta h+\beta mJ(C_{2\sigma-1}-C_{2\sigma}) \right] -\frac{\beta J}{2}(C_{2\sigma-1}-C_{2\sigma})m^2\right \} .
\ee
\end{theorem}
Imposing thermodynamic stability we obtain the following
\begin{proposition}
Even beyond the mean-field level of description, the mixture state of the DHM is described by two independent order parameters, one for each larger cluster, whose self-consistencies read as
\be
m_{1,2}=\tanh(\beta h+\beta Jm_{1,2}(C_{2\sigma-1}-C_{2\sigma})).
\ee
%
\end{proposition}
As for the MF approximation, we are going to find the critical temperature $\beta_c$; considering the system at zero external field $h=0$, thus writing
 \beas
\left\{
\begin{array}{l}
m_1\sim\beta Jm_1(\frac{1}{2^{2\sigma-1} -1} - \frac{1}{2^{2\sigma} -2^{-2\sigma}})+\mathcal{O}(m_1^3),\\\\
m_2\sim\beta Jm_2(\frac{1}{2^{2\sigma-1} -1} - \frac{1}{2^{2\sigma} -2^{-2\sigma}})+\mathcal{O}(m_2^3),\\
\end{array}
\right.
\eeas
so to get the following
\begin{corollary}
This non-mean-field criticality, in the DHM, has the classical exponent too but a different critical temperature $\beta_c^{NMF}$ given by the following formula:
\be
\beta_c^{NMF}=\frac{(2^{2\sigma}-1)(1-2^{1-2\sigma})}{J}.
\ee
It is worth noticing that the non-mean-field interpolation we exploited returned classical (i.e. wrong) critical behavior: this is due to the too rude assumption of self-averaging for the dimers liying in the lowest levels.
\end{corollary}
Comparing the values of $\beta_c^{MF}$ and $\beta_c^{NMF}$, we get the following bound
\be
\beta_c^{NMF}>\beta_c^{MF}\qquad \Rightarrow\qquad T_c^{MF}>T_c^{NMF}.
\ee
We do not push further how analysis here as we want to present a streamlined minimal theory, but the model admits a proliferation of meta-stable states -achievable proceeding hierarchically with the Amit's ansatz -hence taking both the blocks built by $k^k$ spins and splitting them into sub-clusters of $2^{k-1}$ spins each and so on (and correspondingly the hierarchical neural network has a much richer phase diagram w.r.t. its mean-field counterpart): Further investigations can be found in \cite{noiNN}.

\section{Analysis of the Hopfield hierarchical model}

As we saw in the previous Section, the Dyson model has a rich variety of retrievable states, where with {\em retrievable} we mean that they are free energy minima in the thermodynamic limit and their basins of attraction are not negligible. Now we want to apply the analysis previously outlined and the ideas that stemmed from the related findings to a {\em Hierarchical Hopfield Model}.
\newline
To this task we need to introduce, beyond $2^{k+1}$ dichotomic spins/neurons, also $p$ quenched patterns $\bold{\xi}^{\mu}$, $\mu \in (1,...,p)$, that do not participate in thermalization: These are vectors of length $2^{k+1}$, whose entries are extracted once for all from centered and symmetrical i.i.d. as
\be
P(\xi_i^{\mu}) = \frac12 \delta(\xi_i^{\mu}-1)  + \frac12 \delta(\xi_i^{\mu}+1).
\ee

Mirroring the previous Section, the Hamiltonian of the hierarchical Hopfield model is as well defined recursively by the following
\begin{definition}
The Hamiltionian of Hierarchical Hopfield model (HHM) is defined by
\be
H_{k+1}(\vec{S})=H_k(\vec{S_1})+H_k(\vec{S_2})-\frac{1}{2}\frac{1}{2^{2\sigma(k+1)}}\sum_{\mu=1}^{p}\sum_{i,j=1}^{2^{k+1}}\xi^{\mu}_i\xi^{\mu}_jS_iS_j
\ee
with $H_0(S)=0$; $\sigma \in (1/2,1)$ is a number tuning the interaction strength with the neuron's distance, and $p$ is the number of stored patterns. Accounting for the presence of external stimuli can be included simply within a one-body additional term in the Hamiltonian as $\propto h_{\mu}\sum_i^{2^{k+1}}\xi_i^{\mu}\sigma_i$, and a survey overall the stimuli is accomplished summing over $\mu \in (1,...,p)$ all the $h_{\mu}$.
\end{definition}
Even in this context, we can again write the Hamiltonian of the HHM in terms of a distance $d_{i,j}$ between the spin pair $(i,j)$ (see Fig.$1$ panel $B$) obtaining
\begin{eqnarray}
H_{k}[\{S_1...S_{2^k}\}]&=&-\sum_{i<j}S_i S_j\left[ \sum_{l=d_{i,j}}^{k}  \left( \frac{\sum_{\mu=1}^P \xi_i^\mu \xi_j^\mu}{2^{2\sigma l}} \right) \right ]=-\sum_{i<j}S_i S_j \widetilde{J_{ij}},\\
\widetilde{J_{ij}}&=&\sum_{l=d_{i,j}}^{k}\left (\frac{\sum_{\mu=1}^P \xi_i^\mu \xi_j^\mu}{2^{2\sigma l}} \right)=J(d_{i,j},k,\sigma)\sum_{\mu=1}^P \xi_i^\mu \xi_j^\mu
\end{eqnarray}
where, keeping the previous expression (see eq. $3$) to encode neuronal distance, it also holds that
\be
\widetilde{J_{ij}}=\frac{4^{\sigma -d_{i,j}  \sigma }-4^{-k \sigma }}{4^{\sigma }-1}\cdot\sum_{\mu=1}^P \xi_i^\mu \xi_j^\mu,
\ee
hence the Hebbian kernel on a hierarchical topology becomes modified by the distance-dependent weight $J(d_{i,j},k,\sigma)$.
Before starting to implement our interpolative strategy, some definitions are in order.
\begin{definition}
We introduce the Mattis magnetizations (or Mattis overlaps), over the whole system, as
\be
m_{\mu}(\vec{S}) = \frac{1}{2^{k+1}}\sum_{i=1}^{2^{k+1}}\xi_i^{\mu} S_i.
\ee
The definition can be extended trivially to the inner clusters restricting properly the sum over the (pertinent) spins, e.g. dealing with the two larger sub-clusters as before we have
\be
m^{\mu}_{left}=\frac{1}{2^k}\sum_{i=1}^{2^k}\xi_i^{\mu} S_i, \ \ \ \ \ \ %
m^{\mu}_{right}=\frac{1}{2^k}\sum_{j=2^k+1}^{2^{k+1}}\xi_j^{\mu} S_j.
\ee
\end{definition}

\subsection{The thermodynamic limit}

As for the previous investigation, at first we want to prove that the model is well defined, namely that the thermodynamic limit for the free energy exists. To this task we have the following
\begin{theorem}
The thermodynamic limit of the HHM's free energy exists and we call
$$\lim_{k\rightarrow\infty} f_{k+1}(\beta,p,\sigma)=f(\beta,p,\sigma).$$
\end{theorem}
Let us write the Hamiltonian as
$$H_{k+1}(\vec{S})=H_k(\vec{S_1})+H_k(\vec{S_2})-\frac{1}{2}2^{(k+1)}2^{(k+1)(1-2\sigma)}\sum_{\mu=1}^p (m_{\mu}^{k+1}(\vec{S}))^2,$$
and let us consider the following interpolation, where again -for the sake of simplicity- hereafter we stress the dependence by the external fields $ \{ h_{\mu}\}$ only and use the symbol $\mathbb{E}_{\xi}$ to denote averaging over the quenched patterns:
\begin{eqnarray}
&&\Phi_{k+1,t}(\{h_{\mu}\})=\\
\nonumber
&=& \frac{1}{2^{k+1}}\mathbb{E}_{\xi}\log\sum_{\vec{S}}\exp \left \{ \beta \left [ H_k(\vec{S_1})+H_k(\vec{S_2})+t\frac{1}{2}2^{(k+1)}2^{(k+1)(1-2\sigma)}\sum_{\mu=1}^p(m_{\mu}^{k+1}(\vec{S}))^2 +\sum_{\mu=1}^p h_{\mu}\xi^{\mu}_iS_i \right ] \right \} .
\end{eqnarray}
We notice that
\begin{eqnarray}
\Phi_{k+1,1}(h)&=&f_{k+1},\\
\Phi_{k+1,0}(h)&=&f_k
\end{eqnarray}
and that
\be
\frac{d}{dt}\Phi_{k+1,t}=\left\langle \frac{1}{2^{k+1}}\frac{2^{(k+1)}2^{(k+1)(1-2\sigma)}}{2}\sum_{\mu=1}^p [ m_{\mu}^{k+1}(\vec{S})]^2\right\rangle_t\geq 0.\label{f_k_nonIncreasing}
\ee
in such a way that $f_{k+1}(\beta,p,\sigma)\geq f_k(\beta,p,\sigma).$ Now we want to prove that $f_{k+1}(\beta,p,\sigma)$ is bounded: it is enough to see that
\be
f_{k+1}(\beta,p,\sigma)=f_k(\beta,p,\sigma)+\int_0^1\frac{d}{dt}\Phi_{k+1,t}:
\ee
Since we have
\be
\frac{d}{dt}\Phi_{k+1,t}=\left\langle\beta\frac{2^{(k+1)}2^{(k+1)(1-2\sigma)}}{2}\sum_{\mu=1}^p(m_{\mu}^{k+1}(\vec{S}))^2\right\rangle_t\leq  \beta p\frac{2^{(k+1)}2^{(k+1)(1-2\sigma)}}{2},
\ee
we can write
\be
f_{k+1}(\beta,p,\sigma)\leq f_k(\beta,p,\sigma)+\beta p\frac{2^{(k+1)(1-2\sigma)}}{2}.
\ee
Iterating this procedure over the levels we get
\be
f_{k+1}(\beta,p,\sigma)\leq f_0(\beta,p,\sigma)+\frac{\beta p}{2}\sum_{l=1}^{k+1}2^{l(1-2\sigma)},
\ee
such that, in the $k\rightarrow\infty$ limit, we can write
$$f\leq f_0+\frac{\beta p}{2}\sum_{l=1}^{\infty}2^{l(1-2\sigma)}.$$

Since $\sigma>\frac{1}{2}$ the series on the r.h.s. of the above inequality converges, thus $f(\beta,p,\sigma)$ is bounded  by
$$f(\beta,p,\sigma)\leq f_0+\frac{\beta p}{2}\frac{1}{2^{(2\sigma-1)} -1}$$
and non increasing for (\ref{f_k_nonIncreasing}), then its thermodynamic limit exists.\\

\subsection{The mean-field scenario}

Plan of this Section is to investigate the serial and parallel retrieval capabilities in the HHM at the mean-field level. As usual, we obtain our goal by mixing the Amit ansatz technique (in selecting suitably candidate states for retrieval) with the interpolation technique.
\newline
\begin{definition}
Let us define the interpolating Hamiltonian $H_{k+1,t}(\vec{S})$  as
\be\label{crown}
H_{k+1,t}(\vec{S})=H_{k}(\vec{S_1})+H_{k}(\vec{S_2})-\frac{t}{2\cdot2^{2\sigma(k+1)}}\sum_{\mu=1}^p\sum_{i,j=1}^{2^{k+1}}\xi^{\mu}_i\xi^{\mu}_jS_iS_j
-(1-t)\cdot 2^{(k+1)(1-2\sigma)}\sum_{\mu=1}^{p}m_{\mu}\sum_{i=1}^{2^{k+1}}\xi_i^{\mu}S_i,
\ee
\end{definition}
Clearly, we can associate such an Hamiltonian to an extended partition function $Z_{k+1,t}(h)$ and to an extended free energy $\Phi_{k+1,t}(h)$ as
\begin{eqnarray}
Z_{k+1,t}(\{h_{\mu}\})&=&\sum_{\vec{S}}\exp{ \left\{ -\beta \left[ H_{k+1,t}(\vec{S})+\sum_{\mu=1}^{p}h_{\mu}\sum_{i=1}^{2^{k+1}}\xi_i^{\mu}S_i \right] \right \} },\\
\Phi_{k+1,t}(\{h_{\mu}\})&=&\frac{1}{2^{k+1}}\mathbb{E}_{\xi}\log Z_{k+1,t}(\{h_{\mu}\}),
\end{eqnarray}
where, for the sake of simplicity, we stressed only the dependence by the fields.
We can rewrite (\ref{crown}) as
\be
H_{k+1,t}(\vec{S})=H_k(\vec{S_1})+H_k(\vec{S_2})-\frac{2^{2(k+1)}t}{2\cdot2^{2\sigma(k+1)}}\sum_{\mu=1}^p m_{k+1,\mu}^2(\vec{S})-(1-t)2^{(k+1)}2^{(1-2\sigma)(k+1)}\sum_{\mu=1}^p m_{k+1,\mu}(\vec{S}) m_{\mu},
\ee
It is easy to show that
\begin{eqnarray}
\Phi_{k+1,1}(\{h_{\mu}\})&=&f_{k+1},\\
\Phi_{k+1,0}(\{h_{\mu}\})&=&\Phi_{k,1}(\{h_{\mu}+2^{(k+1)(1-2\sigma)}m_{\mu}\}), \label{recursive}
\end{eqnarray}
and that
\beas
\frac{d\Phi_{k+1,t}}{dt}&=&\frac{1}{2^{k+1}}\frac{1}{Z_{k+1,t}}\sum_{\vec{S}}\exp(-\beta( H_{k+1,t}(\vec{S})+\sum_{\mu=1}^p h_{\mu}\sum_{i=1}^{2^{k+1}}\xi_i^{\mu}S_i))(-\beta\frac{dH_{k+1,t}(\vec{S})}{dt})\\
&=&\frac{1}{2^{k+1}}\frac{1}{Z_{k+1,t}}\sum_{\vec{S}}\exp(-\beta (H_{k+1,t}(\vec{S})+\sum_{\mu=1}^p h_{\mu}\sum_{i=1}^{2^{k+1}}\xi_i^{\mu}S_i))\times\\&&\times(\frac{\beta 2^{2(k+1)}}{2\cdot 2^{2\sigma(k+1)}}\sum_{\mu =1}^{p}m_{k+1,\mu}^2(\vec{S})-\beta 2^{(k+1)(1-2\sigma)}2^{(k+1)}\sum_{\mu=1}^{p}m_{\mu}m_{k+1,\mu}(\vec{S}))\\
&=&\frac{\beta}{2}2^{(k+1)(1-2\sigma)}\left\langle \sum_{\mu=1}^{p}m_{k+1,\mu}^2(\vec{S})-2m_{\mu}m_{k+1,\mu}(\vec{S})+m_{\mu}^2\right\rangle_t-\frac{\beta}{2}2^{(k+1)(1-2\sigma)}\sum_{\mu=1}^p m_{\mu}^2\\
&=& \frac{\beta}{2}2^{(k+1)(1-2\sigma)}\sum_{\mu=1}^p\left\langle (m_{\mu}^{k+1}(\vec{S})-m_{\mu})^2\right\rangle_t-\frac{\beta}{2}2^{(k+1)(1-2\sigma)}\sum_{\mu=1}^p m_{\mu}^2.
\eeas
Since the term in the brackets above $\left\langle \cdot\right\rangle_t$ is nonnegative, we get
\beas
\Phi_{k+1,1}&=&\Phi_{k+1,0}+\int_{0}^1\frac{d\Phi_{k+1,t}(x,h)}{dt}dt\\
&\geq &\Phi_{k,1}(\{h_{\mu}+2^{(k+1)(1-2\sigma)}m_{\mu}\})-\frac{\beta}{2}2^{(k+1)(1-2\sigma)}\sum_{\mu=1}^p m_{\mu}^2\\
&\geq & \Phi_{1,0}(\{h_{\mu}+\sum_{l=2}^{k+1}2^{l(1-2\sigma)}m_{\mu}\})-\frac{\beta}{2}\sum_{l=2}^{k+1}2^{l(1-2\sigma)}\sum_{\mu=1}^p m_{\mu}^2 \\
&=&\Phi_{0,1}(\{h_{\mu}+\sum_{l=1}^{k+1}2^{l(1-2\sigma)}m_{\mu}\})-\frac{\beta}{2}\sum_{l=1}^{k+1}2^{l(1-2\sigma)}\sum_{\mu=1}^p m_{\mu}^2,
\eeas
where we used (\ref{recursive}) recursively.
\newline
Now we can estimate the last term,  $\Phi_{0,1}(\{h_{\mu}+\sum_{l=1}^{k+1}2^{l(1-2\sigma)}m_{\mu}\})$, in the following way
\begin{eqnarray}
\Phi_{0,1}(\{h_{\mu}+\sum_{l=1}^{k+1}2^{l(1-2\sigma)}m_{\mu}\})&=&\mathbb{E}_{\xi}\log\sum_{S\in\{-1,1\}}\exp(\beta\sum_{\mu=1}^p(h_{\mu}+\sum_{l=1}^{k+1}2^{l(1-2\sigma)}m_{\mu})\xi^{\mu}S)\\
&=&\log2+\mathbb{E}_{\xi}\log\cosh(\beta\sum_{\mu=1}^p (h_{\mu}+\sum_{l=1}^{k+1}2^{l(1-2\sigma)}m_{\mu})\xi^{\mu}),
\end{eqnarray}
where $\mathbb{E}_{\xi}$ averages over the quenched patterns as usual.
\newline
Summarizing we have
\be
f_{k+1}\geq \log 2+\mathbb{E}_{\xi}\log\cosh(\beta\sum_{\mu=1}^p (h_{\mu}+\sum_{l=1}^{k+1}2^{l(1-2\sigma)}m_{\mu})\xi^{\mu})-\frac{\beta}{2}\sum_{l=1}^{k+1}2^{l(1-2\sigma)}\sum_{\mu=1}^p m_{\mu}^2.
\ee
which is enough to state the next
\begin{theorem} (Mean Field Bound for Serial Retrieval)
Given $-1\leq m_{\mu}\leq +1$, $\forall \mu=1,...,p$ the following relation holds
$$f(\beta,\{h_{\mu}\},p)\geq \sup_{\{m^{\mu}\}}\left[ \log 2+\mathbb{E}_{\xi}\log\cosh(\beta\sum_{\mu=1}^p (h_{\mu}+C_{2\sigma-1}m_{\mu})\xi^{\mu})-\frac{\beta}{2}C_{2\sigma-1}\sum_{\mu=1}^p m_{\mu}^2\right],$$
where the optimal order parameters are the solutions of the system
\be
m^{\mu}=\mathbb{E}_{\xi}\xi^{\mu}\tanh (\beta\sum_{\nu=1}^p(h_{\nu}+C_{2\sigma-1}m^{\nu})\xi^{\nu}),\nonumber
\ee
that are the self-consistent equations of a standard Hopfield model with rescaled temperature $\beta C_{2\sigma-1}$.
\end{theorem}
Again the critical temperature of the model with no external fields, separating the paramagnetic phase from the retrieval one, can be obtained expanding for small $\{m^{\mu}\}$, so to get
\be
m^{\mu}=\mathbb{E}_{\xi}[\beta C_{2\sigma-1}\xi^{\mu}\sum_{\nu=1}^p(\xi^{\nu}m^{\nu})]+ \mathcal{O}({m^{\mu}}^2)=\beta C_{2\sigma-1}+ \mathcal{O}({m^{\mu}}^2)
\ee
hence $\beta^{MF}_c=C_{2\sigma -1}^{-1}$. As previously outlined for the DHM, it is possible to assume -for the $k^{th}$ level- two different classes of Mattis magnetizations $m_{left}^{\mu}=m_{1}^{\mu}$ and $m_{right}^{\mu}=m_{2}^{\mu}$ such that $m^{\mu}=m_1^{\mu}+m_2^{\mu}$ and then check the stability of this potential parallel retrieval of two patterns. Following this way we write
\beas
\Phi_{k,1}(\{h_{\mu}+2^{(k+1)(1-2\sigma)}m^{\mu}\})=\frac{1}{2}\Phi^1_{k,1}(\{h_{\mu}+2^{(k+1)(1-2\sigma)}m^{\mu}\})
+\frac{1}{2}\Phi^2_{k,1}(\{h_{\mu}+2^{(k+1)(1-2\sigma)}m^{\mu}\}).
\eeas
Using the procedure developed in the previous analysis for both the elements of the sum and using, starting from the $k$-th level, $m^{\mu}_{1,2}$ as order parameters of $\Phi^{1,2}$ we obtain
\begin{eqnarray}\nonumber
f_{k+1}&\geq& \frac{1}{2}\Phi_{0,1}(\{h_{\mu}+\sum_{l=1}^{k}2^{l(1-2\sigma)}m_1^{\mu} +2^{(k+1)(1-2\sigma)}m^{\mu}\})
+\frac{1}{2}\Phi_{0,1}(\{h_{\mu}+\sum_{l=1}^{k}2^{l(1-2\sigma)}m_2^{\mu} \\ &+& 2^{(k+1)(1-2\sigma)}m^{\mu}\})
-\frac{\beta}{2}2^{(k+1)(1-2\sigma)}\sum_{\mu=1}^pm_{\mu}^2-\frac{\beta}{2}\sum_{l=1}^k {2^{l(1-2\sigma)}\sum_{\mu=1}^p\frac{(m_1^{\mu})^2+(m_1^{\mu})^2}{2}}.
\end{eqnarray}
Now, evaluating both the terms $\Phi_{0,1}$ and taking the infinite volume limit we can finally state the next
\begin{theorem} (Mean Field Bound for Parallel Retrieval) Given $-1\leq m_{\mu}\leq +1$, $\forall \mu=1,...,p$ the following relation holds
\begin{eqnarray}
f(\beta,\{h_{\mu}\},p)&\geq& \sup_{\{m^{\mu}\}}[ \log 2+\mathbb{E}_{\xi}\log\cosh(\beta\sum_{\mu=1}^p (h_{\mu}+C_{2\sigma-1}m^{\mu}_1)\xi^{\mu})\nonumber \\
&+& \mathbb{E}_{\xi}\log\cosh(\beta\sum_{\mu=1}^p (h_{\mu}+C_{2\sigma-1}m^{\mu}_2)\xi^{\mu})-\frac{\beta}{2}C_{2\sigma-1}\sum_{\mu=1}^p \frac{{m^{\mu}_1}^2+{m^{\mu}_2}^2}{2}],
\end{eqnarray}
representing the free energy  of two effectively independent Hopfield models -one for each subcluster (left and right), whose optimal order parameters fulfill
$$
m^{\mu}_{1,2}=\mathbb{E}_{\xi}\xi^{\mu}\tanh (\beta\sum_{\nu=1}^p(h_{\nu}+C_{2\sigma-1}m^{\nu}_{1,2})\xi^{\nu})
$$
and whose critical temperature is again $\beta_c^{MF}=C_{2\sigma -1}^{-1}$.
\end{theorem}

\subsection{The not-mean-field scenario}

Scope of the present Section is to bypass mean-field limitations and show that the outlined scenario is robust.
To this task, mirroring the previous analysis on DHM, here we provide an improved (with respect to the mean-field counterpart) bound.
\newline
The idea underlying this non-mean-field bound is the same that we used in the DHM, extensively explained in \cite{DH}.
Let us start introducing the following
\begin{definition}
Let us take $x\geq 0$ -a real scalar parameter related to order parameter fluctuations-, and $t \in [0,1]$ -which allows the morphism between the tricky two body coupling and the effective one-body interaction-, and let us introduce also the following interpolating Hamiltonian
\be H_{k+1,t}=-tu(\vec{S})-(1-t)v(\vec{S})+H_k(\vec{S_1})+H_k(\vec{S_2})\label{NMFInterpolation}\ee
with
\begin{eqnarray}
u(\vec{S})&=&\frac{1}{2\cdot2^{2\sigma(k+1)}}\sum_{\mu=1}^p\sum_{i,j}^{2^{k+1}}\xi_i^{\mu}\xi_j^{\mu}S_iS_j+\frac{x}{2\cdot 2^{2\sigma(k+1)}}\sum_{\mu=1}^p\sum_{i,j=1}^{2^{k+1}}(\xi_i^{\mu}S_i-m_{\mu})(\xi_j^{\mu}S_j-m_{\mu}),\\
v(\vec{S})&=&\frac{(x+1)}{2\cdot 2^{2\sigma(k+1)}}(\sum_{\mu=1}^p\sum_{i,j=1}^{2^k}(\xi_i^{\mu}S_i-m_{\mu})
(\xi_j^{\mu}S_j-m_{\mu})+\sum_{i,j=2^k+1}^{2^{k+1}}(\xi_i^{\mu}S_i-m_{\mu})(\xi_j^{\mu}S_j-m_{\mu}))\\
&+&\sum_{\mu=1}^p m_{\mu}2^{(k+1)(1-2\sigma)}\sum_{i=1}^{2^{k+1}}\xi_i^{\mu}S_i.
\end{eqnarray}
\end{definition}
The partition function and free energy associated to the Hamiltonian $(\ref{NMFInterpolation})$ are, respectively,
\begin{eqnarray}
Z_{k+1,t}(x,\{h_{\mu}\})&=&\sum_{\vec{S}}\exp(-\beta (H_{k+1,t}(\vec{S})+\sum_{\mu=1}^p\sum_i^{2^{k+1}}h_i^{\mu}\xi_i^{\mu}S_i)),\label{eqZ}\\
\Phi_{k+1,t}(x,\{h_{\mu}\})&=&\frac{1}{2^{k+1}}\mathbb{E}_{\xi}\log Z_{k+1,t}(x,\{h_{\mu}\}).\label{eqPhi}
\end{eqnarray}
As usual we relate $\Phi_{k+1,0}$ with $\Phi_{k,1}$ as
\be
\Phi_{k+1,0}(x,\{h_{\mu}\})=\Phi_{k,1}(\frac{1+x}{2^{2\sigma}},\{h_{\mu}+ m_{\mu}2^{(k+1)(1-2\sigma)}\}).\label{PhiIterative}
\ee
It is possible to show that the derivative of $\Phi_{k+1,t}$ with respect to $t$ is
\begin{eqnarray}\label{NMFflux}\small\small\nonumber
\frac{d\Phi_{k+1,t}}{dt}(x,t)&=&\frac{1}{2^{k+1}}\frac{1}{Z_{k+1,t}}\sum_{\vec{S}}\exp(-\beta (H_{k+1,t}(\vec{S})+\sum_{\mu=1}^p h_{\mu}\sum_i^{2^{k+1}}\xi_i^{\mu}S_i))(\beta u(\vec{S})-\beta v(\vec{S})) \\
=-\frac{\beta}{2}2^{(k+1)(1-2\sigma)}\sum_{\mu=1}^p m_{\mu}^2&+&\frac{\beta(x+1)}{2^{(k+1)(1+2\sigma)}}
\sum_{\mu=1}^p\sum_{1\leq i\leq 2^k}\sum_{2^k+1\leq j\leq2^{k+1}}\left\langle (\xi_i^{\mu}S_i-m_{\mu})(\xi_j^{\mu}S_j-m_{\mu})\right\rangle_t.
\end{eqnarray}
Now we are going to neglect the fluctuation source, containing $\left\langle (\xi_i^{\mu}S_i-m_{\mu})(\xi_j^{\mu}S_j-m_{\mu})\right\rangle_t$, and that we indicate with $\mathcal{C}(k+1,\beta,\sigma,\{m_{\mu}\})$: at difference with before, while in the pure ferromagnetic case Griffiths inequalities hold \cite{Griffiths1,Griffiths2} and ensure that such a term is positive defined (thus allowing us to get the bound), in this context -as for neural networks Griffiths theory have not yet been developed- we are left with an approximation only. However we stress that this is not a big deal as, already at a mean-field level, while the true solution of the Hopfield model is expected to be full-RSB (see e.g. \cite{JSP-RS,spinglassgauss,bipCWSK,tirozzi}) usually only its replica symmetric approximation is retained for practical purposes (where order parameter's fluctuations are disregarded) and it is indeed an {\em approximation} and not a bound.
\begin{eqnarray}
f_{k+1}=\Phi_{k+1,1}(0,\{h_{\mu}\})&=&\Phi_{k,1}(\frac{1}{2^{2\sigma}},\{h_{\mu}+\beta m_{\mu}2^{(k+1)(1-2\sigma)}\})-\frac{\beta}{2}2^{(k+1)(1-2\sigma)}\sum_{\mu=1}^p m_{\mu}^2\nonumber \\
&+&\mathcal{C}(k+1,\beta,\sigma,\{m_{\mu}\})
\end{eqnarray}
Iterating the procedure one arrives to:
\begin{eqnarray}
f_{k+1}&=& \Phi_{0,1}(\sum_{l=1}^{k+1} 2^{-2l\sigma},\{h_{\mu}+\beta m_{\mu}\sum_{l=1}^{k+1}2^{l(1-2\sigma)}\})-\frac{\beta}{2}\sum_{l=1}^{k+1}2^{l(1-2\sigma)}\sum_{\mu=1}^p m_{\mu}^2\nonumber \\
&+& \sum_{l=1}^{k+1} \mathcal{C}(l,\beta,\sigma,\{m_{\mu}\}).
\end{eqnarray}
Calculating the value of $\Phi_{0,1}$, using the $(\ref{eqZ})$, $(\ref{eqPhi})$ and $(\ref{PhiIterative})$ we get the following
%

\begin{theorem}(Non-mean field approximation for Serial retrieval) Given $-1\leq m_{\mu}\leq +1$, $\forall \mu=1,...,p$ the Serial NMF-approximation for the Hierarchical Hopfield model reads as
$$f^{NMF}(\beta, \{h_{\mu}\},p)= \sup_{m} \left[\log 2+\mathbb{E}_{\xi}\log\cosh(\sum_{\mu=1}^{p} (h_{\mu}+\beta m_{\mu}(C_{2\sigma-1}-C_{2\sigma}))\xi^{\mu})-\frac{\beta}{2}\sum_{\mu=1}^p m_{\mu}^2(C_{2\sigma-1}-C_{2\sigma})\right],$$
representing an Hopfield model at rescaled temperature, with optimal order parameters fulfilling
$$
m^{\mu}=\mathbb{E}_{\xi}\xi^{\mu}\tanh (\beta\sum_{\nu=1}^p(\beta h_{\nu}+(C_{2\sigma-1}-C_{2\sigma})m^{\nu})\xi^{\nu})
$$
and critical temperature $\beta^{NMF}_c=C_{2\sigma-1}-C_{2\sigma}$.
\end{theorem}
Again it is possible to generalize the serial retrieval, assuming two different families of Mattis magnetizations $(\{m^{\mu}_{1,2}\}_{\mu=1}^p)$ for the two blocks of spin under the $k$-th level. Following this way and using the NMF interpolating procedure for the two blocks we get
\begin{eqnarray}
f_{k+1}(\{h_{\mu}\}, \beta, \sigma,p)&=&\log 2+\frac 1 2 \mathbb{E}_{\xi}\log\cosh(\sum_{\mu=1}^{p} (\beta h_{\mu}+\beta m^{\mu}_1(\sum_{l=1}^{k}2^{l(1-2\sigma)}-\sum_{l=1}^{k+1}2^{l(-2\sigma)})+\beta m^{\mu}2^{(k+1)(1-2\sigma)})\xi^{\mu})\nonumber \\
&+&\frac 1 2 \mathbb{E}_{\xi}\log\cosh(\sum_{\mu=1}^{p} (\beta h_{\mu}+\beta m^{\mu}_2(\sum_{l=1}^{k}2^{l(1-2\sigma)}-\sum_{l=1}^{k+1}2^{l(-2\sigma)})+\beta m^{\mu}2^{(k+1)(1-2\sigma)})\xi^{\mu})\nonumber \\
&-&\frac{\beta}{2}(\sum_{l=1}^{k}2^{l(1-2\sigma)}-\sum_{l=1}^{k+1}2^{l(-2\sigma)})\sum_{\mu=1}^p \frac{{m^{\mu}_1}^2+{m^{\mu}_2}^2}{2}
-\frac{\beta}{2}2^{(k+1)(1-2\sigma)}\sum_{\mu=1}^p m_{\mu}^2\nonumber \\
&+& \mathcal{C}(k+1,\beta,\sigma,\{m_{\mu}\})+\frac 1 2 \sum_{l=1}^k\left(\mathcal{C}(l,\beta,\sigma,\{m^1_{\mu}\})+\mathcal{C}(l,\beta,\sigma,\{m^2_{\mu}\})\right)
\end{eqnarray}
that, in the infinite volume limit, where the interactions between the two block vanish, and partially neglecting again the correlations, brings to the following
\begin{definition}
(Non mean field approximation for Parallel retrieval)  Given $-1\leq m_{\mu}\leq +1$, $\forall \mu=1,...,p$ the Parrallel NMF-approximation for the Hierarchical Hopfield model reads as
\begin{eqnarray}
f(\{h_{\mu}\}, \beta, \sigma,p)&=&\sup_{\{m^{\mu}_{1,2}\}} \Big \{ \log 2+\frac 1 2 \mathbb{E}_{\xi}\log\cosh \Big [ \sum_{\mu=1}^{p} (\beta h_{\mu}+\beta m^{\mu}_1(C_{2\sigma-1}-C_{2\sigma}) \Big] \nonumber \\
&+&\frac 1 2 \mathbb{E}_{\xi}\log\cosh \Big[ \sum_{\mu=1}^{p} (\beta h_{\mu}+\beta m^{\mu}_2(C_{2\sigma-1}-C_{2\sigma}) \Big] \nonumber \\
&-&\frac{\beta}{2}(C_{2\sigma-1}-C_{2\sigma})\sum_{\mu=1}^p \frac{{m^{\mu}_1}^2+{m^{\mu}_2}^2}{2} \Big \},
\end{eqnarray}
i.e., the free energy of two independent Hopfield models for each of the two subgroups of spins, with disentangled optimal order parameters satisfying
$$
m^{\mu}_{1,2}=\mathbb{E}_{\xi}\xi^{\mu}\tanh (\beta\sum_{\nu=1}^p(h_{\nu}+(C_{2\sigma-1}-C_{2\sigma})m^{\nu}_{1,2}\xi^{\nu}),
$$
and critical temperature $\beta^{NMF}_c=C_{2\sigma-1}-C_{2\sigma}$.
\end{definition}

\section{Outlooks and conclusions}

Originally, neural networks were developed on fully connected structures and embedded with mean field constraints \cite{hopfield}, later on -as far as graph theory analyzed complex structures as small worlds \cite{corre} or scale free networks \cite{scale}, neural networks have been readily implemented on these structures too \cite{troia,tonno1,tonno2}, hence neurons were no longer fully connected, but the mean-field prescription was retained. Note that in those cases parallel processing was extensive, up to $P\sim N$, but pattern-vectors allowed (extensive) blank entries \cite{PRL}.
\newline
However the quest to bypass mean-field limitation, beyond driven already by clear physical arguments, has been recently strongly emphasized directly from neurobiology, and right toward hierarchical prescriptions \cite{NatComm}.
\newline
As a sideline, recently, exactly hierarchical models experienced a renewed interest in statistical mechanics as structures where testing spin-glasses beyond the mean-field paradigm \cite{REM,giorgione}, thus implicitly offering the backbone for bypassing mean-field limitations in neural networks too \footnote{Note that {\em hierarchical} models in neural networks already appeared \cite{gutfreund,angel,perez} but in those papers the adjective was refereed to the -correlated- patterns and not to the neurons: a completely different research.}.
\newline
As we recently developed a new interpolation scheme for these structures \cite{DH} that, while do not fully solving the model's thermodynamics yet, allows however to overcome the mean-field picture still keeping a formal description (i.e. theorems and bounds available), we extended such a technology to cover neural networks too and we mixed it with the Amit technique of investigating by an Ansatz the candidate retrievable states: this fusion resulted in a stronger method that allowed to analyze both the ferromagnet on a hierarchical topology (Dyson hierarchical model) as well as the neural network on a hierarchical topology (Hopfield hierarchical model).
\newline
Starting with the former (that we used as a test-guide for the latter), remarkably, beyond the ferromagnetic scenario already largely discussed \cite{Dyson,Gallavotti,Sinai}, we have shown that the model has a huge plethora of meta-stable states that become stable in the thermodynamic limit and forbid self-averaging for the magnetization in a way quite similar to the scenario deserved for the overlap in mean-field spin-glasses \cite{broken,MPV}.
\newline
Filtering these results within the neural network perspective, we have been able to show that these networks -where clusters of neurons well far apart essentially do not interact- perform both a' la Hopfield \cite{hopfield}, hence relaxing via a global rearrangement of all the spins in order to retrieve an extensive stored pattern and in a multitasking fashion very close to the parallel processing performance shown by other (mean-field) associative networks developed by us in the past two years in a series of paper \cite{PRL,Sollich,ton1,ton2,PRE}.
\newline
A last note of interest, regards the capacity of these networks: we have shown how it is possible to recall simultaneously two patterns by spitting the system into two subgroups, going down over the levels from the top and we have seen that, since the upper interaction is vanishing with enough velocity (see further Appendix A on this point), in the thermodynamic limit the two subgroups of neurons can be thought of as {\em independent}: each one is governed by an Hopfield Hamiltonian and can choose to recall one of the memorized patterns. Clearly we could use the same argument iteratively and split the system in more sub-sub-clusters going down over the various levels. Crucially, what is fundamental is that -at least- the sum of the upper levels of interactions remains vanishing in the infinite volume limit. If we split the system $M$ times, we have to use different order parameters, for the magnetizations of the blocks, until the $k-M$ level, where the system is divided into $2^M$ subgroups. The procedure keeps working as far  as
\be
\lim_{k\to\infty} \sum_{l=k-M}^k 2^{l(1-2\sigma)} \sum_{\mu=1}^p m^{\mu}_{l}=0.
\ee
Since the magnetizations are bounded, in the worst case we have
\begin{eqnarray}
\sum_{l=k-M}^k 2^{l(1-2\sigma)} \sum_{\mu=1}^p m^{\mu}_{l}&\leq& p \sum_{l=k-M}^k 2^{l(1-2\sigma)} \nonumber\\
&\leq& p \sum_{l=k-M}^\infty 2^{l(1-2\sigma)} \propto  2^{(1-2\sigma)(k-M)} p:
\end{eqnarray}
if we want the system to handle up to  $p$ patterns, we need $p$ different blocks of spins and then $M=\log(p)$. So for example if $p=\mathcal{O}(k)$, $2^{(1-2\sigma)(k-\log(p))} p \to 0$ as $k\to\infty$.
\newline
Thus the parallel processing ability works at best with a logarithmic load of patterns, as far as $p=\mathcal{O}(k)=\mathcal{O}(\log N)$, however such a bound -which is however not enormous as the serial counterpart handle up to $P \sim N$ patterns- is never reach in practice:  why do these networks have a restricted capacity?
\newline
The answer lies in the real fingerprint of not-mean-field spontaneous parallel processing: in order to explain the network ability to manage two patterns contemporarily, we used the argument that the upper links connecting the two communities {\em left} and {\em right} are actually vanishing in the thermodynamic limit. While this is exciting for parallel processing capabilities as it allows to divide the network into almost-disjoint communities, is however a disaster for the storage capacity as each time we can use this argument, we are effectively admitting that a huge amount of synapses for storing the memories are vanishing. We found thus a novel balancing requirement in non-mean-field processing: extreme parallel processing implies smallest storage capacity and viceversa: we aim to check this prescription on real networks in the future.

\section*{Acknowledgments}
The Authors are grateful to Sapienza Universit\`{a} di Roma, Gruppo Nazionale per la Fisica Matematica (GNFM-INdAM) and Istituto Nazionale di Fisica Nucleare (INFN) for supporting their work.
\newline
EA and AB are indebted to Enzo Marinari and Federico Ricci-Tersenghi for a list of countless items, ranging from personal support to priceless discussions.

\section*{Appendix A: Selection of a state}
While we have shown that, in the thermodynamic limit, the (intensive) energies associated to the two states that we used as example (the ferromagnetic and the mixture states) do coincide, thus thermodynamically the metastable state is not forbidden (while its weight is negligible w.r.t. the ferromagnetic scenario, and the system must be trapped opportunely with external fields in its basin to keep it in the large $k$ limit), we still have to face the following addressable question: Let us consider the mixture state and approach the critical region from the ergodic scenario: the two clusters differ in magnetization, one has $m_1>0$ and the other $m_1<0$ and there is the just the upper (hence weakest) link connecting them. Maybe that one cluster acts on the other playing as an external field in mean-field schemes, thus selecting the phase /reversing the other cluster magnetization sign), which would result in destruction of mixture states?  Aim of this note is to show that this is not the case.
\newline
The way we pave to prove this statement is the following: at first we will address this question within the more familiar mean-field perspective (namely considering the Curie-Weiss model), then we will enlarge the observation stemmed in that example toward bipartite ferromagnetic systems and we will show that they continue holding. As a last step to obtain the result, we will compare the Dyson model (whose spins are locked in a mixture state) with a bipartite ferromagnet so to enlarge to the present model the stability argument.

\begin{figure}[tb] \begin{center}
\includegraphics[width=.8\textwidth]{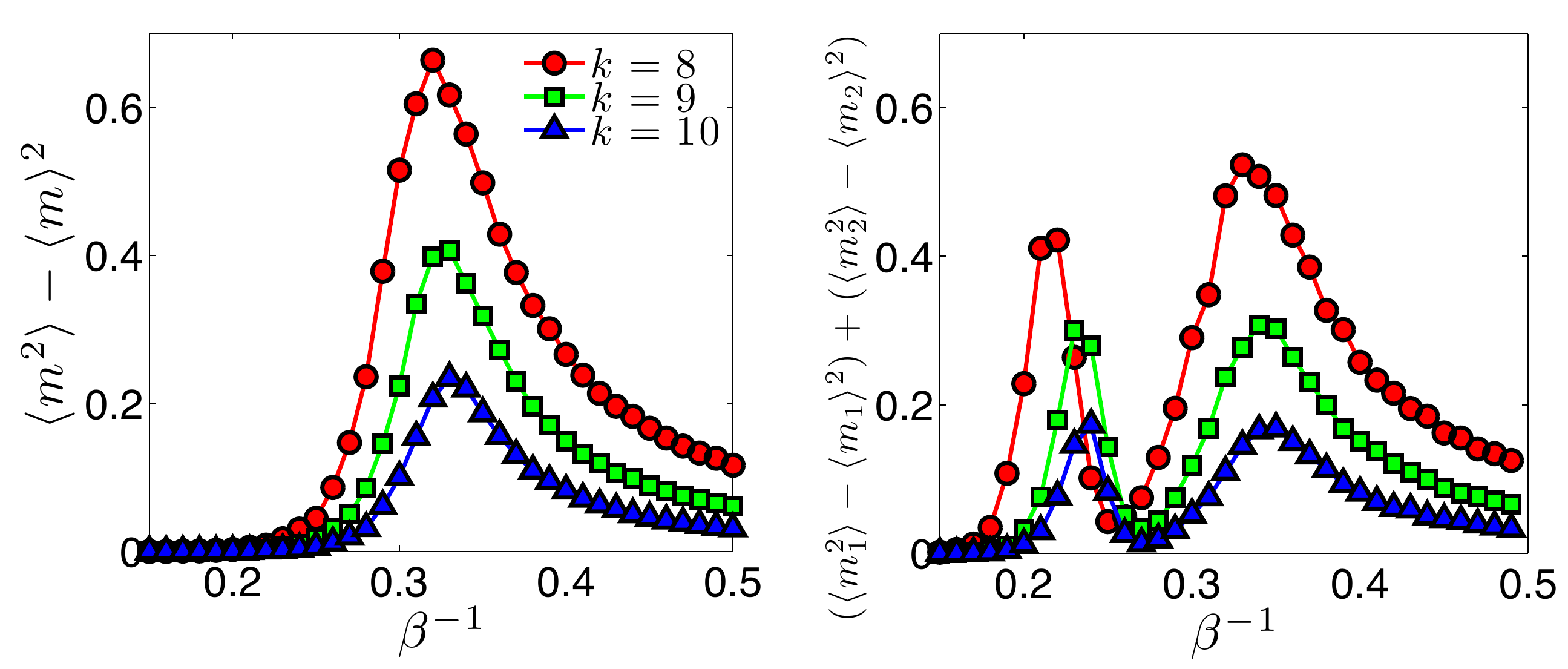}
\caption{Analysis of the susceptibility of the system, defined as $X= \langle m^2 \rangle - \langle m \rangle^2$, versus the noise level $T \equiv \beta^{-1}$, for various sizes (as reported in the legend) and $\sigma=0.99$. Left panel: $X(T)$ for the pure state. Right panel: $X(T)$ for the mixture state. Note that, while in the ferromagnetic (pure) case all the cuspids are on the same noise level whatever $k$, this is not the case for the mixture state because such a state is metastable as the difference in the energy $\Delta E$ among the two states scales as $\propto 1/N^{2\sigma-1}$, hence only for $k\to\infty$ the mixture state becomes stable and its cuspid happens at the same noise level of the pure counterpart.}
\end{center}
\end{figure}

Under the critical temperature systems of spins whose dynamics is no longer ergodic have an equilibrium state (in the thermodynamic limit) that can be a mixture of several pure states. Each of these states has its own basin of attraction in the sense that the system will reach one of them, according with its initial configuration. As far as ferromagnetic systems are concerned, adding a suitable external field, it is possible to select one of this pure states, i.e. the dynamics is forced into one of the attractors. Now we can ask when an external field is able to select a state or not. Consider a Glauber dynamics for a ferromagnetic system of $N$ spins at zero temperature:
\be
S_i(t+1)=\operatorname{sgn}(h_i(S (t)) + h_N).
\ee
For example we can keep in mind the case of the CW model where $h_i(S (t)$, the field acting on the $i$-th spin is the magnetization $m(S(t))=\frac 1 N \sum_{i=1}^N S_i(t)$. In that case each initial configuration for which $|h_N|>|h_i(S^0)|$ will follow the external field. In general (if for example $|h_N|<1$) there will always exist initial configuration ($|h_N|<|h_i(S^0)|$) that do not feel the influence of the external field, but, if we choose the initial configuration randomly and accordingly to $P_N(S^0)$, we can say that the field $h_N$ selects the state if
\be
P_N(S^0 : |h_N|>|h_i(S^0)|)\stackrel{N\to\infty}{\rightarrow}1.
\ee
On the contrary we will say that the field will not select the state if
\be
P_N(S^0 : |h_N|<|h_i(S^0)|)\stackrel{N\to\infty}{\rightarrow}1.
\ee
In what follow we will consider $P_N(S^0)=\prod_{i=1}^N p(S_i^0)$, with $p(S)$ uniform in $\{-1,1\}$. For the CW model we can state the following
\begin{theorem}
In the CW model, where $h_i(S)=m(S)=\frac 1 N\sum_{i=1}^N S_i$, $\forall \epsilon >0$,
\item $h_N : |h_N|> \frac{1}{N^{\frac 1 2 (1-\epsilon)}}$ selects the state;
\item $h_N : |h_N|< \frac{1}{N^{\frac 1 2 (1+\epsilon)}}$ does not select the state.
\end{theorem}
For what concerns the first statement we note that, if $|h_N|> \frac{1}{N^{\frac 1 2 (1-\epsilon)}}$
\begin{eqnarray}
P_N(S^0 : |h_N|>|h_i(S^0)|)&=&1-P_N\left(S^0 : |h_i(S^0)|>|h_N|\right)\geq 1-P_N\left(S^0: |m(S^0)|> \frac{1}{N^{\frac 1 2 (1-\epsilon)}} \right)\nonumber \\
&\geq& 1- N^{1-\epsilon}\mathbb{E}_N[m^2(S^0)]=1-N^{-\epsilon}\stackrel{N\to\infty}{\rightarrow}1,
\end{eqnarray}
where we used Chebyshev inequality and the fact that $\mathbb{E}_N[m^2(S^0)]=\frac 1 N$. For the second statement we note that, since
$|h_N|< \frac{1}{N^{\frac 1 2 (1+\epsilon)}}$,
\begin{eqnarray}
P_N(S^0 : |h_N|>|h_i(S^0)|)&\leq& P_N\left(S^0 : |m(S^0)|<\frac{1}{N^{\frac 1 2 (1+\epsilon)}}\right)\nonumber \\
&=& P_N\left(S^0 : |\sqrt{N}m(S^0)|<\frac{1}{N^{\frac{\epsilon}{2}}}\right)\nonumber\\
&\to& \mu_{\mathcal{N}(0,1)}\left(|z|< \frac{1}{N^{\frac{\epsilon}{2}}}\right) \stackrel{N\to\infty}{\rightarrow} 0,
\end{eqnarray}
where we just used the fact that the variable $\sqrt{N}m(S^0)=\frac 1 {\sqrt{N}} \sum_{i=1}^N S^0_i$ satysfies the CLT and tends in distribution to a $\mathcal{N}(0,1)$ gaussian variable.
\newline
\newline
We can repeat the same analysis in a mean field bipartite ferromagnetic model where the interaction inside the parties (modulated by $J_{11}$ and $J_{22}$) and the ones among the parties (modulated by $J_{12}$) have different couplings. In that case we can ask in which case $J_{12}$ is able to select the state where the two parties are aligned and not independent. If we consider for example a spin in the first party we have for the Glauber dynamics
\be
S_i(t+1)=\operatorname{sgn}(h_i(S (t)) + h_N)=\operatorname{sgn}(J_{11}m_1(S(t)) + J_{12}m_2(S(t))),
\ee
i.e. we can repeat the same argument of the CW model identifying the field sent by the second party (proportional to the magnetization $m_2(S)$) as the external field. Thus, using the analogous version of the previous theorem, we see that $J_{12}$ is able to select the state only if
\be
J_{12}(N)|m_2(S^0)|> \frac{1}{N^{\frac 1 2 (1-\epsilon)}} ,
\ee
with probability $1$. Since for the CLT $|m_2(S^0)|$ is $\mathcal{O}(\sqrt{N})$ with probability one, vanishing $J_{12}$ will not be able to select the totally magnetized state: in that case the system behaves exactly as two non interacting CW subsystems.
The hierarchical model can be considered from this point of view a generalization of a bipartite model. In fact if we divide the system into two subgroups of spins we have that the external field (representing the last level of interaction) is proportional to $J(N)m_N(S)$, while the internal field is a sum of contributions coming from all the submagnetizations. Since $J(N)=N^{1-2\sigma}$ is vanishing in the thermodynamic limit, the two subgroups behave as they were non interacting: this may puzzle about the phase transition as the system -when not trapped within the pure state- crossing the critical line (in the $\beta, \sigma$ plane) moves from an ergodic region -where the global magnetization is zero- toward a mixture state where again is zero. However, regarding the latter, the two sub-clusters have not-zero magnetizations and even in this case, crossing the line returns in a canonical phase transition (see Fig.$3$). To give further proof of this delicate way of breaking ergodicity, we show further results from extensive Monte Carlo runs that confirm our scenario and are reported in Fig.$3$.

\end{document}